\documentclass[reprint,twocolumn,amssymb, nobibnotes, aps,prl,longbibliography]{revtex4-2}
\usepackage{color}
\usepackage{graphicx}
\usepackage{xspace}
\usepackage{nicefrac}
\usepackage[utf8x]{inputenc}

\newcommand{\note}[1]{{\color{black}#1}}
\usepackage{ae,lmodern}
\usepackage[T1]{fontenc}

\begin{document}

\title{\note{Ultra high strength and plasticity mechanisms of Si and SiC} nanoparticles revealed by first principles molecular dynamics}%

\author{L. Pizzagalli}
\email{laurent.pizzagalli@univ-poitiers.fr}
\author{J. Godet}
\affiliation{Institut Pprime, CNRS UPR 3346, Universit\'{e} de Poitiers, Poitiers, France}

\date{}%

\begin{abstract}
It is now well established that materials are stronger when their dimensions are reduced to submicron scale. \note{However,  what happens at dimensions such as a few tens of nanometers or lower remains largely unknown, with conflicting reports on strength or plasticity mechanisms. Here, we combined first principles molecular dynamics and classical force fields to investigate the mechanical properties of 1--2~nm Si and SiC nanoparticles}. These compression simulations unambiguously reveal that the strength continues to increase down to such sizes, and that in these systems the theoretical bulk strength can be reached \note{or even exceeded} in some cases. Most of the nanoparticles yield by amorphization \note{at strains} greater than 20\%, \note{with no evidence of $\beta$-tin phase for Si}. Original and unexpected mechanisms are also identified, such as the homogeneous formation of a dislocation loop embryo for the $\langle111\rangle$ compression of SiC nanoparticles, and an elastic softening for the $\langle001\rangle$ compression of Si nanoparticles. 
\end{abstract}

\maketitle

\note{Exploring materials at the nanoscale is a successful story with the discovery of new properties and original phenomena in various domains like phononics, plasmonics, photonics, etc... In the specific case of nanomechanics, pioneering works by Taylor and Brenner revealed the surprisingly high strength of various metals at the micron scale~\cite{Tay24PR,Bre56JAP}. Later  investigations at lower dimensions~\cite{Won97SCI,Uch04SCI} confirmed the important finding that the reduction of dimensions considerably increases the strength, with values converging towards the theoretical limit~\cite{Zhu09MRSB}.


These works raise fundamental and unresolved questions. First, one may wonder whether strength keeps increasing down to the smallest possible sizes. It has been tentatively proposed that there is a threshold size below which the strength appears to be constant~\cite{Han15ADM,Wag15AM,Sha18NATCOM}. However other studies reported a weakening at sizes equal to a few tens of nm~\cite{Fan09JNR,Chr11NATNAN}. A second unknown concerns the maximum strength value that can be attained at small scale. It is assumed that the intrinsic strength of the perfect bulk crystal represents an upper bound. Values close to this theoretical limit were recently reported for several kind of metal nanoparticles (NP)~\cite{Han15ADM,Sha18NATCOM,Sha20AM}. However, the maximum strength for even smaller NP made of other materials like Si was determined to be significantly lower~\cite{Wag15AM}. These conflicting results clearly call for investigations of strength at the lowest possible scales. 

Other unknown factors are plasticity mechanisms and how they are impacted at low dimensions. It is notorious that materials with covalent bonding become ductile at low dimensions~\cite{Ger12JMR}. But our understanding of the underlying plastic deformation mechanisms remains limited and controversial. In silicon, the most studied covalent material, classical molecular dynamics calculations (MD) predicted either the heterogeneous nucleation of dislocations~\cite{Hal12CMS,Chr11NATNAN,Kil18AM}, or the occurrence of phase transition~\cite{Val07PRL,Hal11CMS} during the compression of 10--50~nm NP. This controversy is possibly caused by classical potentials, which are unable to provide an accurate description of small and highly strained NP~\cite{Amo21CRP}. Experiments are scarce and the most comprehensive study suggests that both phase transition and dislocations could concurrently occur~\cite{Wag15AM}. At the smallest dimensions, i.e. down to a few nm, the situation is even more complicated since amorphization becomes another competitive deformation mode~\cite{Nie07SPR}. Further investigations are needed to address these issues.

Current experimental apparatuses allow for studying the plastic deformation of nanostructures with dimensions as small as a few tens of nm~\cite{Lah14TL,Hin16SM}. But it remains difficult to apply a deformation in a controlled manner at lower sizes. As previously mentioned, the use of classical MD is questionable because of the inaccuracy of interatomic potentials. In order to circumvent this issue and obtain key answers concerning the mechanical properties of NP at the smallest scales, we applied a recently developed approach combining first principles MD together with planar repulsive force fields~\cite{Piz20PRB}, thus allowing for dynamic compression at finite temperature with first principles accuracy~\cite{Piz22DRM,Piz22PCCP}. This approach is used in the present work to investigate the mechanical properties of small Si and SiC NP, aiming at answering fundamental and unresolved questions regarding the strength and plasticity mechanisms in nanometer-scale materials. Our calculations reveal that the strength of NP reaches or even exceeds the bulk theoretical limit, which is an original feat to our knowledge. Incidentally,  we also demonstrate that the strength keeps increasing down to a few nm in the case of Si and SiC, at odds with previous reports~\cite{Fan09JNR,Chr11NATNAN}. We find that these NP yield mostly by amorphization, with no transition to a high pressure crystalline phase. Finally, our simulations also reveal the unexpected homogeneous formation of a small dislocation embryo in SiC NP. Yet such a mechanism was customarily acknowledged to be unfavorable, due to the reduced dimensions or with respect to heterogeneous nucleation.  
}

Two Si and four SiC \note{NP models, with cuboctahedron or truncated octahedron shapes} and with sizes ranging from 1.1~nm to 1.8~nm, are studied~\cite{Sup1}. Car-Parrinello molecular dynamics (CPMD) calculations are performed using the Quantum Espresso code~\cite{Gia17JPCM}. All NP are located in supercells with dimensions adjusted to ensure a minimum vacuum distance of 10~\AA\ between replicas. The electronic structure is computed using a 25~Ry plane wave cutoff, $\gamma$-point sampling, the PBE exchange correlation functional~\cite{Per96PRL}, and ultrasoft pseudopotentials~\cite{Van90PRB}. Other parameters specific to the CPMD method are similar to the ones used in Ref.~\cite{Piz22DRM}. The CPMD timestep is 0.2~fs and the compression speed is 0.1~\AA/ps ensuring a reasonable strain rate~\cite{Amo21CRP} but at the cost of typically $3\times10^5$ ionic iterations.  Compression is done at 300~K in a controlled strain mode along the $\langle111\rangle$ or the $\langle001\rangle$ crystallographic orientations. 

\begin{figure}[th!]
\begin{center}
\includegraphics[width=8.6 cm]{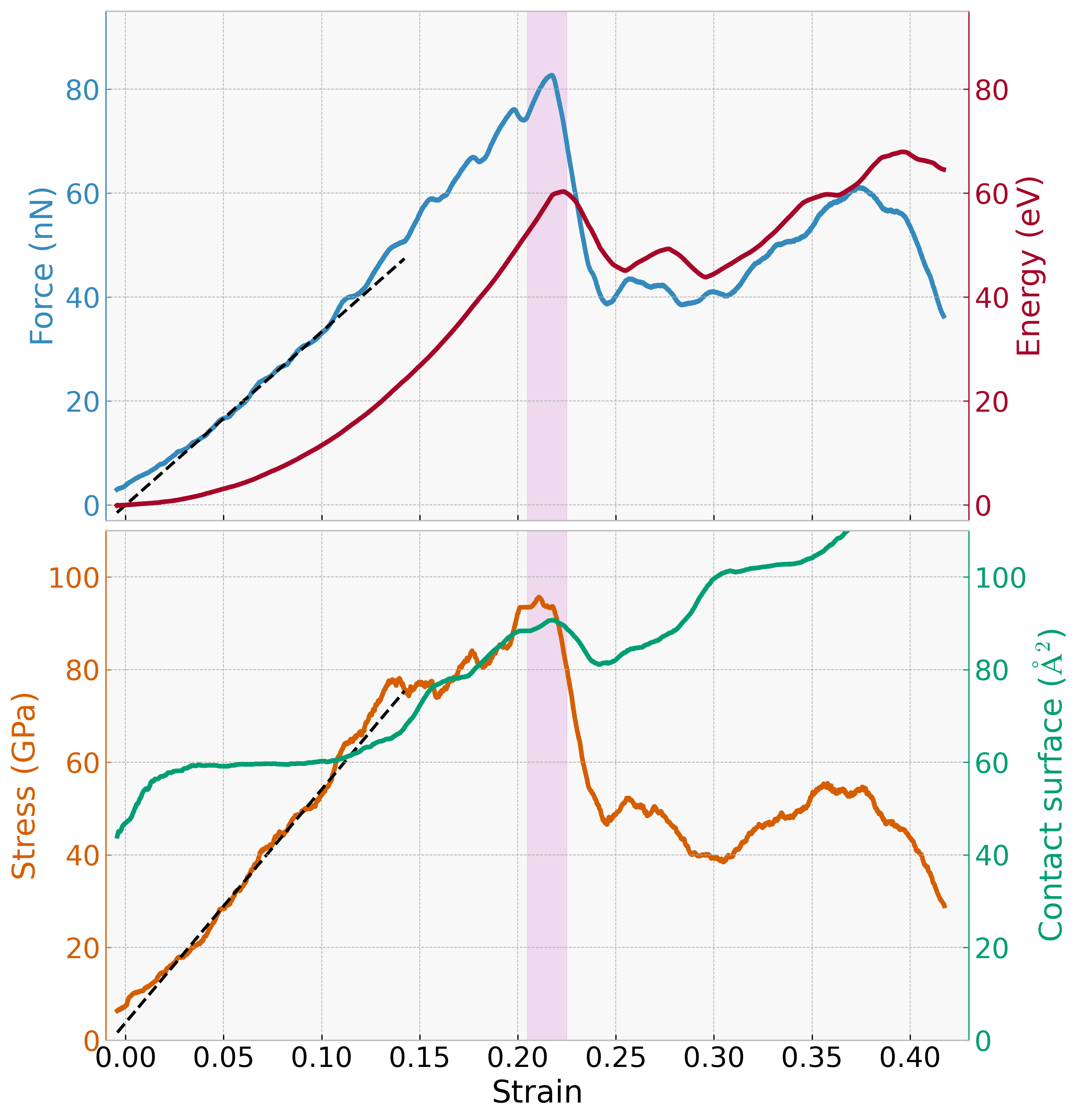}
\caption{Force (nN, blue), energy (eV, red), stress (GPa, orange), and contact surface (\AA$^2$, green) as a function of  strain, for a Si$_{79}$C$_{68}$ NP compressed along $\langle111\rangle$. The displayed contact surface and stress are obtained using the $S_1$ surface definition (see Supp.~Inf.). Dashed lines are linear interpolation of force and stress in the strain range 0.05--0.1 where contact surface is constant. The pink strip shows the strain region where plastic deformation starts.} 
\label{fig:1}
\end{center}
\end{figure}

We first analyze how properties like energy and force applied on the NP vary as a function of compression strain $\varepsilon$. Both are directly obtained from the simulation. The energy averaged before compression is used as the energy reference. The zero strain is estimated from a linear extrapolation of the contact force in the elastic regime, with an accuracy at best of $\pm0.01$. We also determine the contact stress as the ratio of the force to the contact surface. The latter is inherently prone to uncertainty, a fortiori in small systems. We used different calculation methods~\cite{Sup2} to determine a meaningful range of possible contact surface area. Accordingly in the following all stress-related quantities are given as a range of values. 

Figure~\ref{fig:1} shows results for Si$_{79}$C$_{68}$ as an example. At low strains, energy and force exhibit a quadratic and linear variation, respectively, as expected for an elastic deformation. The contact surface is constant in the 0.04--0.11 strain range, leading to an almost linear increase of contact stress. The elastic regime is characterized by a stiffness of 303~N/m, and an elastic modulus of 411--780~GPa (depending on the surface calculation method) \note{in good agreement with the $\langle111\rangle$ bulk SiC value of 541~GPa}. The data and curves for all systems are included in Supplementary Material~\cite{Sup3}. It is found that elastic moduli of NP are close to the corresponding bulk value (for a given material and orientation). As expected, larger moduli are obtained for SiC and $\langle111\rangle$ than for Si and $\langle001\rangle$. For $\varepsilon>0.10$, small ripples are observed in the force curves, in association with weak variations of contact surfaces, while the energies grow smoothly. Similar observations are made for all studied systems. 

\begin{figure}[th!]
\begin{center}
\includegraphics[width=8.6 cm]{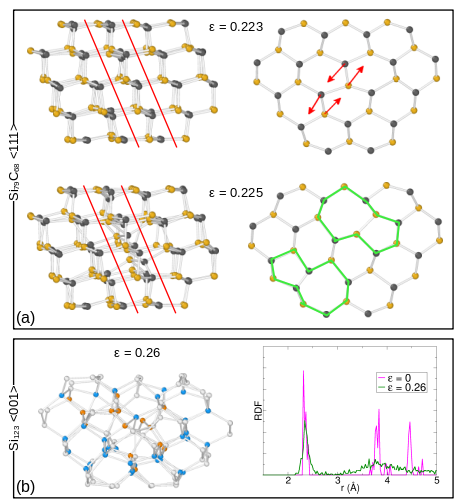}
\caption{Plasticity mechanisms occurring in the Si$_{79}$C$_{68}$ NP compressed along $\langle111\rangle$ (a), and in the Si$_{123}$ NP compressed along $\langle001\rangle$ (b). In (a), pictures on the left show the NP structure at compression strains of 0.223 and 0.225, the compression axis being the vertical of the figure (Si in gold and C in black). Pictures on the right show a flat view of atoms contained in a $\langle111\rangle$-oriented slice (limited by the two red lines in the left pictures). Red arrows indicate the main atomic displacements leading to the formation of a dislocation dipole (emphasized by the green thick lines). In (b), the left picture displays the NP structure at a strain of 0.26, with atoms colored according to a PTM analysis~\cite{Mah16MSMSE} (blue: cubic diamond, orange: hexagonal diamond, white: unidentified structure), whereas the right graph compares RDF for uncompressed and compressed NP. } 
\label{fig:2}
\end{center}
\end{figure}

For the system showcased in Fig.~\ref{fig:1}, the force, energy and stress exhibit large drops when the strain exceeds 0.21--0.23, suggesting the initiation of plastic relaxation.  Figure~\ref{fig:2}-a shows the compressed NP just before ($\varepsilon=0.223$) and after ($\varepsilon=0.225$) the activation of the first plasticity mechanism. The main structural changes concern atoms in a $\langle111\rangle$-normal strip, bordered by the two red lines in the figure. We observe that the stress relaxation mechanism essentially consists of the concerted motion of four atoms along $\langle011\rangle$, Si and C atoms moving in opposite directions (red arrows in the right side of Fig.~\ref{fig:2}-a). It leads to the formation of a point defect dipole (green lines in  Fig.~\ref{fig:2}-a). Interestingly, this configuration is also equivalent to a dislocation core dipole embryo  with a minimal expansion of the dislocation loop. In fact, each core defect exhibits the 5/7-ring structure typical of a $1/2\langle110\rangle$ 60$^\circ$ glide dislocation core~\cite{Rod17AM}. An atomic displacement analysis reveals that the Burgers vector magnitude is about 80\% of the value expected for a dislocation in bulk SiC. This is coherent with previous analyses of the homogeneous nucleation of a dislocation loop at low dimensions~\cite{Mil08JMPS,Gut08AM}. At $\varepsilon=0.242$ one of these cores glides in the Burgers vector direction and annihilates itself at the NP surface, which further confirms the dislocation nature of the defect. The homogeneous nucleation of a dislocation loop in such a small volume was unexpected~\cite{Val07PRL} and never reported to our knowledge, and thus constitutes one of the highlights of the present study. A similar mechanism is also identified for Si$_{61}$C$_{61}$, which suggests it is specific to the $\langle111\rangle$ compression of SiC NP.  

\begin{table}[h!]
\begin{center}
\caption{Identified mechanisms and corresponding strains and stresses (GPa) during the compression.}\label{Tab:1}
\begin{ruledtabular}
\begin{tabular}{lccl}
 Model &   Strain & Stress (GPa)      & Mechanism \\ \hline
 Si$_{123}$ $\langle001\rangle$           &  0.2  & 19--43   & softening \\ 
                                          &  0.26 & 10.7--24   & amorphization \\[.1cm] 
 Si$_{148}$ $\langle111\rangle$           & 0.195 & 20.6--39   & amorphization \\[.1cm]  
 Si$_{71}$C$_{56}$ $\langle001\rangle$    & 0.30  & 64--107   & amorphization \\[.1cm]  
 Si$_{80}$C$_{92}$ $\langle001\rangle$    & 0.24  & 76--170  & NP rotation   \\
                                          & 0.30  & 64--116   & amorphization \\[.1cm] 
 Si$_{61}$C$_{61}$ $\langle111\rangle$    & 0.25  & 85--146  & dislocation embryo \\
                                          & 0.33  & 47--82   & amorphization \\[.1cm] 
 Si$_{79}$C$_{68}$ $\langle111\rangle$    & 0.21  & 80--135  & dislocation embryo   \\
                                          & 0.37  & 49--72   & amorphization  \\ 
\end{tabular}
\end{ruledtabular}
\end{center}
\end{table}

The structural changes associated with large stress drops in all compressed NP were all thoroughly examined and the results reported in Tab.~\ref{Tab:1}. In most cases, Polyhedral Template Matching (PTM) and Radial Distribution Functions (RDF) analyses reveal that a partial crystal-amorphous transition occurs as primary or secondary stress relaxation mechanism (Fig.~\ref{fig:2}-b). This is in agreement with the generally accepted notion of an improved stability of disordered phases at the nanoscale~\cite{Nie07SPR}. Two other events are noteworthy. First, we observe a force/stress maximum at 0.20 strain for the $\langle001\rangle$ compression of Si$_{123}$, with only a small inflection of the energy curve (Tab.~\ref{Tab:1} and Fig.~S9). Atomic and electronic structure analyses, detailed in Supplementary Material, were carried on, which indicate a relatively homogeneous deformation and the absence of amorphous or $\beta$-tin phases formation~\cite{Sup4}. The important finding is that the onset of this $\langle001\rangle$ softening would appear as the ultimate yield stress of the NP by examining the stress--strain curve, although the deformation remains purely elastic, with the first irreversible event occurring at $\varepsilon=0.26$.  The second interesting event is observed during the $\langle001\rangle$ compression of Si$_{80}$C$_{92}$, at $\varepsilon=0.24$. Compared to the previous case, the stress reaches a maximum but not the force or the energy (Fig.~S7 in Supp.~Inf.). The stress drop is due to an increase of contact surface, which is a consequence of the disorientation of the NP relatively to the compression axis (Fig.~S12 in Supp. Inf.).




\begin{figure}[th!]
\begin{center}
\includegraphics[width=8.6 cm]{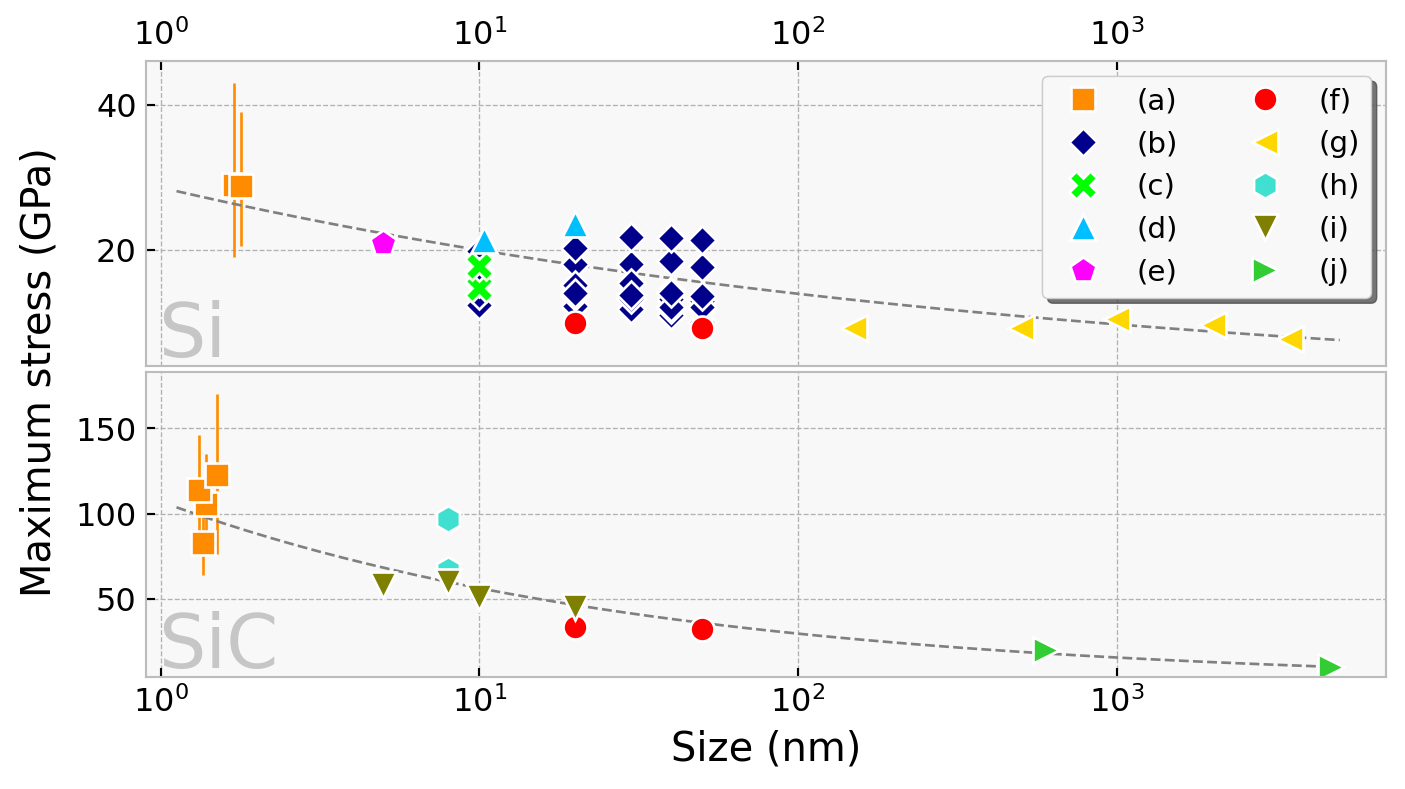}
\caption{Maximum stress (GPa) versus size (nm) plots for Si (top) and SiC (bottom), with (a) data from Tab.~\ref{Tab:1}, or compiled from the literature (b)~\cite{Kil18AM}, (c)~\cite{Hal11CMS}, (d)~\cite{Chr11NATNAN}, (e)~\cite{Val07PRL}, (f)~\cite{Kil19AM}, (g)~\cite{Che20NATCOM}, (h)~\cite{He16JNR}, (i)~\cite{Kay21APA}, (j)~\cite{Shi12JACS}. The orange squares show the averages of the four stress values (depending on surface definition) for each NP. The orange lines run from the lowest to the highest of these stress values for each NP. The dashed grey lines show a power fit to the data (see text for details). }
\label{fig:3}
\end{center}
\end{figure}

We now focus on the highest contact \note{stress $\sigma$ achieved} during compression. In the investigated strain range, \note{$\sigma$ is associated with primary events as reported in Tab.~\ref{Tab:1}}. It is well acknowledged that decreasing dimensions leads to \note{an increase of strength for sizes down to a few tens of nm~\cite{Amo21CRP}}. Conflicting propositions were made for smaller dimensions, with either constant~\cite{Han15ADM,Wag15AM} or  decreasing~\cite{Fan09JNR} strength when size is reduced. Our maximum stress values are represented in Fig.~\ref{fig:3}, together with data from the literature for other Si and SiC systems. Considering the ranges due to surface area definitions, it is clear that our values are overall significantly greater than those computed or measured for larger systems. \note{It suggests that for Si and SiC, and also likely for other zinc-blende or wurtzite materials, the strength keeps increasing down to  a few nm.} Fitting all data with a power law expression $\beta d^{-\alpha}$, $d$ being the size (Fig.~\ref{fig:3}), $\alpha$ exponents of 0.16 (0.28) are obtained for Si (SiC). The Si value is close to  0.08--0.11, the estimated exponents for much larger nanopillars~\cite{Che20NATCOM}, and in the expected range for ceramic-like materials~\cite{Kor11PM}. Note that there is a debate about the value and meaning of these scaling exponents~\cite{Dun13IJP,Fer16AM,Fai19PRM}.

\begin{figure}[th!]
\begin{center}
\includegraphics[width=8.6 cm]{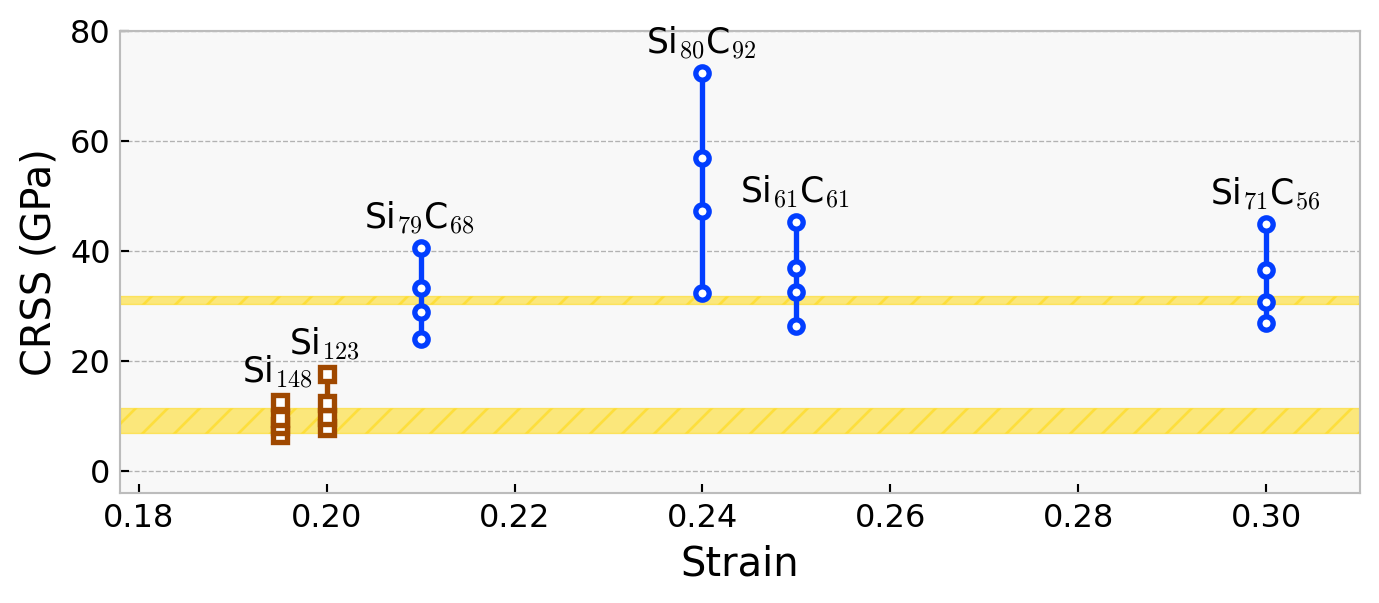}
\caption{Critical resolved shear stress (GPa) versus strain at the first contact stress maximum (brown squares for Si and blue spheres for SiC). For each system, the different CRSS values correspond to the stress values associated with each surface definition. The \note{lower and upper} yellow strips show the ranges of theoretical shear strength, built from extreme values reported for  Si~\cite{Rou01PRB,Dub06PRB} and SiC~\cite{Oga04PRB,Ume08PRB}, respectively.}
\label{fig:4}
\end{center}
\end{figure}

Assuming an homogeneous deformation inside the NP during compression, we determined the critical resolved shear stress (CRSS) \note{$\tau$, which is the projected shear stress for a given slip plane and direction and is obtained using $\tau=m\sigma$, with $m$ the Schmid factor (see Supp. Inf.). $m$ is computed from atomic positions at each compression step for all systems, for a $\langle110\rangle$ slip in \{111\} planes}. Initial values are 0.408 and  0.272 for the $\langle001\rangle$ and $\langle111\rangle$ compression orientations, respectively. At strains associated with the maximum contact stresses, \note{$m$} slightly increases up to 0.41--0.42 for $\langle001\rangle$ and 0.30--0.32 for $\langle111\rangle$. The corresponding CRSS data are shown in Fig.~\ref{fig:4}. As expected, CRSS \note{values} are greater for SiC than for Si. Also, they are relatively close for all clusters, except for Si$_{80}$C$_{92}$ which appears as an outlier. Next we compare our computed values with \note{theoretical shear strength data (TSS) from the literature} (Fig.~\ref{fig:4}). \note{TSS, also called ideal shear stress, is defined as the maximum stress in a perfect crystal under a uniform shear deformation}. Assuming the upper limit for surface areas ($S_3$ in Supp. Inf.), i.e. considering the lowest strength values for each case, we find that the computed CRSS are close or only slightly lower than the TSS. This suggests that in 1--2~nm NP the theoretical shear strength is almost reached. This is even more remarkable since our calculations are carried out at 300~K while TSS is a 0~K calculated quantity. Now considering the other surface definitions described in Supp. Inf., the CRSS values \note{are greater than the TSS} in several cases. This surprising finding might be explained by the large compression strains, since it is known that normal stresses can significantly change the TSS~\cite{Ume08PRB}. Another rationale is that our original assumption is not correct i.e., the stress inside the NP \note{is not homogeneous and then locally not equal to the projection of the compression stress.} \note{Finally, quantum confinement effects might also play a role at small sizes}. 

Overall our CPMD simulations reveal a rich and unexpected picture of the mechanical properties of 1-2~nm Si and SiC NP. Very high strain/stress are required to reach the plastic regime, which is in most cases initiated by amorphization. We also identified the homogeneous formation of a seemingly dislocation embryo as first plastic event. This finding conflicts with the broad consensus that homogeneous dislocation nucleation is prevented in such small volumes. Another interesting and original finding is the occurrence  of a softening along the $\langle001\rangle$ orientation in Si associated with a compression stress maximum, although no phase formation occurs and the deformation remains elastic. The maximum stress values suggest that the NP strengths, computed at 300~K, are close to the theoretical strength of the ideal bulk material at 0~K. It also confirms that for covalent materials the strength keeps increasing when dimensions are reduced to a few nm. \note{Note that in this work we investigated $\langle001\rangle$ and $\langle111\rangle$ compression orientations, which correspond to the normals of well defined facets in Wulff-like NP. Compression along other orientations like $\langle110\rangle$ is likely to lead to NP rotation and reorientation of the compression axis along $\langle001\rangle$ or $\langle111\rangle$.} To conclude, this study opens the way towards a better understanding of mechanical properties at a previously inaccessible scale. The originality and importance of our findings is a clear incentive to apply an equivalent approach to other class of materials, like FCC and BCC metals. 

\begin{acknowledgments}
The computer time for this work was provided by several sources: the Spin Center at the University of Poitiers, the MCIA (Mésocentre de Calcul Intensif Aquitain), and GENCI-CINES (Grant 2020-A0090912035). This work pertains to the French Government program “Investissements d’Avenir” (EUR INTREE, reference ANR-18-EURE-0010, and LABEX INTERACTIFS, reference ANR-11-LABX-0017-01). 
\end{acknowledgments}


\begin{thebibliography}{56}%
\makeatletter
\providecommand \@ifxundefined [1]{%
 \@ifx{#1\undefined}
}%
\providecommand \@ifnum [1]{%
 \ifnum #1\expandafter \@firstoftwo
 \else \expandafter \@secondoftwo
 \fi
}%
\providecommand \@ifx [1]{%
 \ifx #1\expandafter \@firstoftwo
 \else \expandafter \@secondoftwo
 \fi
}%
\providecommand \natexlab [1]{#1}%
\providecommand \enquote  [1]{``#1''}%
\providecommand \bibnamefont  [1]{#1}%
\providecommand \bibfnamefont [1]{#1}%
\providecommand \citenamefont [1]{#1}%
\providecommand \href@noop [0]{\@secondoftwo}%
\providecommand \href [0]{\begingroup \@sanitize@url \@href}%
\providecommand \@href[1]{\@@startlink{#1}\@@href}%
\providecommand \@@href[1]{\endgroup#1\@@endlink}%
\providecommand \@sanitize@url [0]{\catcode `\\12\catcode `\$12\catcode
  `\&12\catcode `\#12\catcode `\^12\catcode `\_12\catcode `\%12\relax}%
\providecommand \@@startlink[1]{}%
\providecommand \@@endlink[0]{}%
\providecommand \url  [0]{\begingroup\@sanitize@url \@url }%
\providecommand \@url [1]{\endgroup\@href {#1}{\urlprefix }}%
\providecommand \urlprefix  [0]{URL }%
\providecommand \Eprint [0]{\href }%
\providecommand \doibase [0]{https://doi.org/}%
\providecommand \selectlanguage [0]{\@gobble}%
\providecommand \bibinfo  [0]{\@secondoftwo}%
\providecommand \bibfield  [0]{\@secondoftwo}%
\providecommand \translation [1]{[#1]}%
\providecommand \BibitemOpen [0]{}%
\providecommand \bibitemStop [0]{}%
\providecommand \bibitemNoStop [0]{.\EOS\space}%
\providecommand \EOS [0]{\spacefactor3000\relax}%
\providecommand \BibitemShut  [1]{\csname bibitem#1\endcsname}%
\let\auto@bib@innerbib\@empty
\bibitem [{\citenamefont {Taylor}(1924)}]{Tay24PR}%
  \BibitemOpen
  \bibfield  {author} {\bibinfo {author} {\bibfnamefont {G.~F.}\ \bibnamefont
  {Taylor}},\ }\href {https://doi.org/10.1103/physrev.23.655} {\bibfield
  {journal} {\bibinfo  {journal} {Phys. Rev.}\ }\textbf {\bibinfo {volume}
  {23}},\ \bibinfo {pages} {655} (\bibinfo {year} {1924})}\BibitemShut
  {NoStop}%
\bibitem [{\citenamefont {Brenner}(1956)}]{Bre56JAP}%
  \BibitemOpen
  \bibfield  {author} {\bibinfo {author} {\bibfnamefont {S.~S.}\ \bibnamefont
  {Brenner}},\ }\href {https://doi.org/10.1063/1.1722294} {\bibfield  {journal}
  {\bibinfo  {journal} {J. Appl. Phys.}\ }\textbf {\bibinfo {volume} {27}},\
  \bibinfo {pages} {1484} (\bibinfo {year} {1956})}\BibitemShut {NoStop}%
\bibitem [{\citenamefont {Wong}\ \emph {et~al.}(1997)\citenamefont {Wong},
  \citenamefont {Sheehan},\ and\ \citenamefont {Lieber}}]{Won97SCI}%
  \BibitemOpen
  \bibfield  {author} {\bibinfo {author} {\bibfnamefont {E.~W.}\ \bibnamefont
  {Wong}}, \bibinfo {author} {\bibfnamefont {P.~E.}\ \bibnamefont {Sheehan}},\
  and\ \bibinfo {author} {\bibfnamefont {C.~M.}\ \bibnamefont {Lieber}},\
  }\href {https://doi.org/10.1126/science.277.5334.1971} {\bibfield  {journal}
  {\bibinfo  {journal} {Science}\ }\textbf {\bibinfo {volume} {277}},\ \bibinfo
  {pages} {1971} (\bibinfo {year} {1997})}\BibitemShut {NoStop}%
\bibitem [{\citenamefont {Uchic}\ \emph {et~al.}(2004)\citenamefont {Uchic},
  \citenamefont {Dinmiduk}, \citenamefont {Florando},\ and\ \citenamefont
  {Nix}}]{Uch04SCI}%
  \BibitemOpen
  \bibfield  {author} {\bibinfo {author} {\bibfnamefont {M.~D.}\ \bibnamefont
  {Uchic}}, \bibinfo {author} {\bibfnamefont {D.~M.}\ \bibnamefont {Dinmiduk}},
  \bibinfo {author} {\bibfnamefont {J.~N.}\ \bibnamefont {Florando}},\ and\
  \bibinfo {author} {\bibfnamefont {W.~D.}\ \bibnamefont {Nix}},\ }\href@noop
  {} {\bibfield  {journal} {\bibinfo  {journal} {Science}\ }\textbf {\bibinfo
  {volume} {305}},\ \bibinfo {pages} {986} (\bibinfo {year}
  {2004})}\BibitemShut {NoStop}%
\bibitem [{\citenamefont {Zhu}\ \emph {et~al.}(2009)\citenamefont {Zhu},
  \citenamefont {Li}, \citenamefont {Ogata},\ and\ \citenamefont
  {Yip}}]{Zhu09MRSB}%
  \BibitemOpen
  \bibfield  {author} {\bibinfo {author} {\bibfnamefont {T.}~\bibnamefont
  {Zhu}}, \bibinfo {author} {\bibfnamefont {J.}~\bibnamefont {Li}}, \bibinfo
  {author} {\bibfnamefont {S.}~\bibnamefont {Ogata}},\ and\ \bibinfo {author}
  {\bibfnamefont {S.}~\bibnamefont {Yip}},\ }\href@noop {} {\bibfield
  {journal} {\bibinfo  {journal} {Mater. Res. Soc. Bull.}\ }\textbf {\bibinfo
  {volume} {34}},\ \bibinfo {pages} {167} (\bibinfo {year} {2009})}\BibitemShut
  {NoStop}%
\bibitem [{\citenamefont {Han}\ \emph {et~al.}(2015)\citenamefont {Han},
  \citenamefont {Huang}, \citenamefont {Ogata}, \citenamefont {Kimizuka},
  \citenamefont {Yang}, \citenamefont {Weinberger}, \citenamefont {Li},
  \citenamefont {Liu}, \citenamefont {Zhang}, \citenamefont {Li}, \citenamefont
  {Ma},\ and\ \citenamefont {Shan}}]{Han15ADM}%
  \BibitemOpen
  \bibfield  {author} {\bibinfo {author} {\bibfnamefont {W.-Z.}\ \bibnamefont
  {Han}}, \bibinfo {author} {\bibfnamefont {L.}~\bibnamefont {Huang}}, \bibinfo
  {author} {\bibfnamefont {S.}~\bibnamefont {Ogata}}, \bibinfo {author}
  {\bibfnamefont {H.}~\bibnamefont {Kimizuka}}, \bibinfo {author}
  {\bibfnamefont {Z.-C.}\ \bibnamefont {Yang}}, \bibinfo {author}
  {\bibfnamefont {C.}~\bibnamefont {Weinberger}}, \bibinfo {author}
  {\bibfnamefont {Q.-J.}\ \bibnamefont {Li}}, \bibinfo {author} {\bibfnamefont
  {B.-Y.}\ \bibnamefont {Liu}}, \bibinfo {author} {\bibfnamefont {X.-X.}\
  \bibnamefont {Zhang}}, \bibinfo {author} {\bibfnamefont {J.}~\bibnamefont
  {Li}}, \bibinfo {author} {\bibfnamefont {E.}~\bibnamefont {Ma}},\ and\
  \bibinfo {author} {\bibfnamefont {Z.-W.}\ \bibnamefont {Shan}},\ }\href
  {https://doi.org/10.1002/adma.201500377} {\bibfield  {journal} {\bibinfo
  {journal} {Adv. Mater.}\ }\textbf {\bibinfo {volume} {27}},\ \bibinfo {pages}
  {3385} (\bibinfo {year} {2015})}\BibitemShut {NoStop}%
\bibitem [{\citenamefont {Wagner}\ \emph {et~al.}(2015)\citenamefont {Wagner},
  \citenamefont {Hintsala}, \citenamefont {Kumar}, \citenamefont {Gerberich},\
  and\ \citenamefont {Mkhoyan}}]{Wag15AM}%
  \BibitemOpen
  \bibfield  {author} {\bibinfo {author} {\bibfnamefont {A.~J.}\ \bibnamefont
  {Wagner}}, \bibinfo {author} {\bibfnamefont {E.~D.}\ \bibnamefont
  {Hintsala}}, \bibinfo {author} {\bibfnamefont {P.}~\bibnamefont {Kumar}},
  \bibinfo {author} {\bibfnamefont {W.~W.}\ \bibnamefont {Gerberich}},\ and\
  \bibinfo {author} {\bibfnamefont {K.~A.}\ \bibnamefont {Mkhoyan}},\ }\href
  {https://doi.org/https://doi.org/10.1016/j.actamat.2015.08.029} {\bibfield
  {journal} {\bibinfo  {journal} {Acta Mater.}\ }\textbf {\bibinfo {volume}
  {100}},\ \bibinfo {pages} {256 } (\bibinfo {year} {2015})}\BibitemShut
  {NoStop}%
\bibitem [{\citenamefont {Sharma}\ \emph {et~al.}(2018)\citenamefont {Sharma},
  \citenamefont {Hickman}, \citenamefont {Gazit}, \citenamefont {Rabkin},\ and\
  \citenamefont {Mishin}}]{Sha18NATCOM}%
  \BibitemOpen
  \bibfield  {author} {\bibinfo {author} {\bibfnamefont {A.}~\bibnamefont
  {Sharma}}, \bibinfo {author} {\bibfnamefont {J.}~\bibnamefont {Hickman}},
  \bibinfo {author} {\bibfnamefont {N.}~\bibnamefont {Gazit}}, \bibinfo
  {author} {\bibfnamefont {E.}~\bibnamefont {Rabkin}},\ and\ \bibinfo {author}
  {\bibfnamefont {Y.}~\bibnamefont {Mishin}},\ }\href@noop {} {\bibfield
  {journal} {\bibinfo  {journal} {Nature Communications}\ }\textbf {\bibinfo
  {volume} {9}},\ \bibinfo {pages} {4102} (\bibinfo {year} {2018})}\BibitemShut
  {NoStop}%
\bibitem [{\citenamefont {Fang}\ \emph {et~al.}(2009)\citenamefont {Fang},
  \citenamefont {Weng},\ and\ \citenamefont {Ju}}]{Fan09JNR}%
  \BibitemOpen
  \bibfield  {author} {\bibinfo {author} {\bibfnamefont {K.-C.}\ \bibnamefont
  {Fang}}, \bibinfo {author} {\bibfnamefont {C.-I.}\ \bibnamefont {Weng}},\
  and\ \bibinfo {author} {\bibfnamefont {S.-P.}\ \bibnamefont {Ju}},\ }\href
  {https://doi.org/10.1007/s11051-008-9396-x} {\bibfield  {journal} {\bibinfo
  {journal} {J. Nanopart. Res.}\ }\textbf {\bibinfo {volume} {11}},\ \bibinfo
  {pages} {581} (\bibinfo {year} {2009})}\BibitemShut {NoStop}%
\bibitem [{\citenamefont {Chrobak}\ \emph {et~al.}(2011)\citenamefont
  {Chrobak}, \citenamefont {Tymiak}, \citenamefont {Beaber}, \citenamefont
  {Ugurlu}, \citenamefont {Gerberich},\ and\ \citenamefont
  {Nowak}}]{Chr11NATNAN}%
  \BibitemOpen
  \bibfield  {author} {\bibinfo {author} {\bibfnamefont {D.}~\bibnamefont
  {Chrobak}}, \bibinfo {author} {\bibfnamefont {N.}~\bibnamefont {Tymiak}},
  \bibinfo {author} {\bibfnamefont {A.}~\bibnamefont {Beaber}}, \bibinfo
  {author} {\bibfnamefont {O.}~\bibnamefont {Ugurlu}}, \bibinfo {author}
  {\bibfnamefont {W.}~\bibnamefont {Gerberich}},\ and\ \bibinfo {author}
  {\bibfnamefont {R.}~\bibnamefont {Nowak}},\ }\href@noop {} {\bibfield
  {journal} {\bibinfo  {journal} {Nature Nanotechnology}\ }\textbf {\bibinfo
  {volume} {6}},\ \bibinfo {pages} {480} (\bibinfo {year} {2011})}\BibitemShut
  {NoStop}%
\bibitem [{\citenamefont {Sharma}\ \emph {et~al.}(2020)\citenamefont {Sharma},
  \citenamefont {Kositski}, \citenamefont {Kovalenko}, \citenamefont
  {Mordehai},\ and\ \citenamefont {Rabkin}}]{Sha20AM}%
  \BibitemOpen
  \bibfield  {author} {\bibinfo {author} {\bibfnamefont {A.}~\bibnamefont
  {Sharma}}, \bibinfo {author} {\bibfnamefont {R.}~\bibnamefont {Kositski}},
  \bibinfo {author} {\bibfnamefont {O.}~\bibnamefont {Kovalenko}}, \bibinfo
  {author} {\bibfnamefont {D.}~\bibnamefont {Mordehai}},\ and\ \bibinfo
  {author} {\bibfnamefont {E.}~\bibnamefont {Rabkin}},\ }\href
  {https://doi.org/10.1016/j.actamat.2020.07.054} {\bibfield  {journal}
  {\bibinfo  {journal} {Acta Mater.}\ }\textbf {\bibinfo {volume} {198}},\
  \bibinfo {pages} {72} (\bibinfo {year} {2020})}\BibitemShut {NoStop}%
\bibitem [{\citenamefont {Gerberich}\ \emph {et~al.}(2012)\citenamefont
  {Gerberich}, \citenamefont {Stauffer}, \citenamefont {Beaber},\ and\
  \citenamefont {Tymiak}}]{Ger12JMR}%
  \BibitemOpen
  \bibfield  {author} {\bibinfo {author} {\bibfnamefont {W.~W.}\ \bibnamefont
  {Gerberich}}, \bibinfo {author} {\bibfnamefont {D.~D.}\ \bibnamefont
  {Stauffer}}, \bibinfo {author} {\bibfnamefont {A.~R.}\ \bibnamefont
  {Beaber}},\ and\ \bibinfo {author} {\bibfnamefont {N.~I.}\ \bibnamefont
  {Tymiak}},\ }\href@noop {} {\bibfield  {journal} {\bibinfo  {journal} {J.
  Mater. Research}\ }\textbf {\bibinfo {volume} {27}},\ \bibinfo {pages} {552}
  (\bibinfo {year} {2012})}\BibitemShut {NoStop}%
\bibitem [{\citenamefont {Hale}\ \emph {et~al.}(2012)\citenamefont {Hale},
  \citenamefont {Zhang}, \citenamefont {Zhou}, \citenamefont {Zimmerman},
  \citenamefont {Moody}, \citenamefont {Dumitrica}, \citenamefont {Ballarini},\
  and\ \citenamefont {Gerberich}}]{Hal12CMS}%
  \BibitemOpen
  \bibfield  {author} {\bibinfo {author} {\bibfnamefont {L.}~\bibnamefont
  {Hale}}, \bibinfo {author} {\bibfnamefont {D.-B.}\ \bibnamefont {Zhang}},
  \bibinfo {author} {\bibfnamefont {X.}~\bibnamefont {Zhou}}, \bibinfo {author}
  {\bibfnamefont {J.}~\bibnamefont {Zimmerman}}, \bibinfo {author}
  {\bibfnamefont {N.}~\bibnamefont {Moody}}, \bibinfo {author} {\bibfnamefont
  {T.}~\bibnamefont {Dumitrica}}, \bibinfo {author} {\bibfnamefont
  {R.}~\bibnamefont {Ballarini}},\ and\ \bibinfo {author} {\bibfnamefont
  {W.}~\bibnamefont {Gerberich}},\ }\href@noop {} {\bibfield  {journal}
  {\bibinfo  {journal} {Comp. Mat. Sci.}\ }\textbf {\bibinfo {volume} {54}},\
  \bibinfo {pages} {280 } (\bibinfo {year} {2012})}\BibitemShut {NoStop}%
\bibitem [{\citenamefont {Kilymis}\ \emph {et~al.}(2018)\citenamefont
  {Kilymis}, \citenamefont {G{\'{e}}rard}, \citenamefont {Amodeo},
  \citenamefont {Waghmare},\ and\ \citenamefont {Pizzagalli}}]{Kil18AM}%
  \BibitemOpen
  \bibfield  {author} {\bibinfo {author} {\bibfnamefont {D.}~\bibnamefont
  {Kilymis}}, \bibinfo {author} {\bibfnamefont {C.}~\bibnamefont
  {G{\'{e}}rard}}, \bibinfo {author} {\bibfnamefont {J.}~\bibnamefont
  {Amodeo}}, \bibinfo {author} {\bibfnamefont {U.}~\bibnamefont {Waghmare}},\
  and\ \bibinfo {author} {\bibfnamefont {L.}~\bibnamefont {Pizzagalli}},\
  }\href {https://doi.org/10.1016/j.actamat.2018.07.063} {\bibfield  {journal}
  {\bibinfo  {journal} {Acta Mater.}\ }\textbf {\bibinfo {volume} {158}},\
  \bibinfo {pages} {155} (\bibinfo {year} {2018})}\BibitemShut {NoStop}%
\bibitem [{\citenamefont {Valentini}\ \emph {et~al.}(2007)\citenamefont
  {Valentini}, \citenamefont {Gerberich},\ and\ \citenamefont
  {Dumitric\ifmmode~\u{a}\else \u{a}\fi{}}}]{Val07PRL}%
  \BibitemOpen
  \bibfield  {author} {\bibinfo {author} {\bibfnamefont {P.}~\bibnamefont
  {Valentini}}, \bibinfo {author} {\bibfnamefont {W.~W.}\ \bibnamefont
  {Gerberich}},\ and\ \bibinfo {author} {\bibfnamefont {T.}~\bibnamefont
  {Dumitric\ifmmode~\u{a}\else \u{a}\fi{}}},\ }\href
  {https://doi.org/10.1103/PhysRevLett.99.175701} {\bibfield  {journal}
  {\bibinfo  {journal} {Phys. Rev. Lett.}\ }\textbf {\bibinfo {volume} {99}},\
  \bibinfo {pages} {175701} (\bibinfo {year} {2007})}\BibitemShut {NoStop}%
\bibitem [{\citenamefont {Hale}\ \emph {et~al.}(2011)\citenamefont {Hale},
  \citenamefont {Zhou}, \citenamefont {Zimmerman}, \citenamefont {Moody},
  \citenamefont {Ballarini},\ and\ \citenamefont {Gerberich}}]{Hal11CMS}%
  \BibitemOpen
  \bibfield  {author} {\bibinfo {author} {\bibfnamefont {L.}~\bibnamefont
  {Hale}}, \bibinfo {author} {\bibfnamefont {X.}~\bibnamefont {Zhou}}, \bibinfo
  {author} {\bibfnamefont {J.}~\bibnamefont {Zimmerman}}, \bibinfo {author}
  {\bibfnamefont {N.}~\bibnamefont {Moody}}, \bibinfo {author} {\bibfnamefont
  {R.}~\bibnamefont {Ballarini}},\ and\ \bibinfo {author} {\bibfnamefont
  {W.}~\bibnamefont {Gerberich}},\ }\href
  {https://doi.org/http://dx.doi.org/10.1016/j.commatsci.2010.12.023}
  {\bibfield  {journal} {\bibinfo  {journal} {Comp. Mat. Sci.}\ }\textbf
  {\bibinfo {volume} {50}},\ \bibinfo {pages} {1651 } (\bibinfo {year}
  {2011})}\BibitemShut {NoStop}%
\bibitem [{\citenamefont {Amodeo}\ and\ \citenamefont
  {Pizzagalli}(2021)}]{Amo21CRP}%
  \BibitemOpen
  \bibfield  {author} {\bibinfo {author} {\bibfnamefont {J.}~\bibnamefont
  {Amodeo}}\ and\ \bibinfo {author} {\bibfnamefont {L.}~\bibnamefont
  {Pizzagalli}},\ }\href {https://doi.org/10.5802/crphys.70} {\bibfield
  {journal} {\bibinfo  {journal} {Comptes Rendus Physique}\ }\textbf {\bibinfo
  {volume} {22}},\ \bibinfo {pages} {1} (\bibinfo {year} {2021})}\BibitemShut
  {NoStop}%
\bibitem [{\citenamefont {Ni{\`{e}}pce}\ and\ \citenamefont
  {Pizzagalli}(2007)}]{Nie07SPR}%
  \BibitemOpen
  \bibfield  {author} {\bibinfo {author} {\bibfnamefont {J.~C.}\ \bibnamefont
  {Ni{\`{e}}pce}}\ and\ \bibinfo {author} {\bibfnamefont {L.}~\bibnamefont
  {Pizzagalli}},\ }in\ \href {https://doi.org/10.1007/978-3-540-72993-8_2}
  {\emph {\bibinfo {booktitle} {Nanomaterials and Nanochemistry}}}\ (\bibinfo
  {publisher} {Springer Berlin Heidelberg},\ \bibinfo {year} {2007})\ pp.\
  \bibinfo {pages} {35--54}\BibitemShut {NoStop}%
\bibitem [{\citenamefont {Lahouij}\ \emph {et~al.}(2014)\citenamefont
  {Lahouij}, \citenamefont {Dassenoy}, \citenamefont {Vacher}, \citenamefont
  {Sinha}, \citenamefont {Brass},\ and\ \citenamefont {Devine}}]{Lah14TL}%
  \BibitemOpen
  \bibfield  {author} {\bibinfo {author} {\bibfnamefont {I.}~\bibnamefont
  {Lahouij}}, \bibinfo {author} {\bibfnamefont {F.}~\bibnamefont {Dassenoy}},
  \bibinfo {author} {\bibfnamefont {B.}~\bibnamefont {Vacher}}, \bibinfo
  {author} {\bibfnamefont {K.}~\bibnamefont {Sinha}}, \bibinfo {author}
  {\bibfnamefont {D.~A.}\ \bibnamefont {Brass}},\ and\ \bibinfo {author}
  {\bibfnamefont {M.}~\bibnamefont {Devine}},\ }\href
  {https://doi.org/10.1007/s11249-013-0246-3} {\bibfield  {journal} {\bibinfo
  {journal} {Tribol. Lett.}\ }\textbf {\bibinfo {volume} {53}},\ \bibinfo
  {pages} {91} (\bibinfo {year} {2014})}\BibitemShut {NoStop}%
\bibitem [{\citenamefont {Hintsala}\ \emph {et~al.}(2016)\citenamefont
  {Hintsala}, \citenamefont {Wagner}, \citenamefont {Gerberich},\ and\
  \citenamefont {Mkhoyan}}]{Hin16SM}%
  \BibitemOpen
  \bibfield  {author} {\bibinfo {author} {\bibfnamefont {E.}~\bibnamefont
  {Hintsala}}, \bibinfo {author} {\bibfnamefont {A.}~\bibnamefont {Wagner}},
  \bibinfo {author} {\bibfnamefont {W.}~\bibnamefont {Gerberich}},\ and\
  \bibinfo {author} {\bibfnamefont {K.}~\bibnamefont {Mkhoyan}},\ }\href
  {https://doi.org/http://dx.doi.org/10.1016/j.scriptamat.2015.12.004}
  {\bibfield  {journal} {\bibinfo  {journal} {Scripta Materialia}\ }\textbf
  {\bibinfo {volume} {114}},\ \bibinfo {pages} {51 } (\bibinfo {year}
  {2016})}\BibitemShut {NoStop}%
\bibitem [{\citenamefont {Pizzagalli}(2020)}]{Piz20PRB}%
  \BibitemOpen
  \bibfield  {author} {\bibinfo {author} {\bibfnamefont {L.}~\bibnamefont
  {Pizzagalli}},\ }\href@noop {} {\bibfield  {journal} {\bibinfo  {journal}
  {Phys. Rev. B}\ }\textbf {\bibinfo {volume} {102}},\ \bibinfo {pages}
  {094102} (\bibinfo {year} {2020})}\BibitemShut {NoStop}%
\bibitem [{\citenamefont {Pizzagalli}(2022{\natexlab{a}})}]{Piz22DRM}%
  \BibitemOpen
  \bibfield  {author} {\bibinfo {author} {\bibfnamefont {L.}~\bibnamefont
  {Pizzagalli}},\ }\href {https://doi.org/10.1016/j.diamond.2022.108870}
  {\bibfield  {journal} {\bibinfo  {journal} {Diamond \& Related Materials}\
  }\textbf {\bibinfo {volume} {123}},\ \bibinfo {pages} {108870} (\bibinfo
  {year} {2022}{\natexlab{a}})}\BibitemShut {NoStop}%
\bibitem [{\citenamefont {Pizzagalli}(2022{\natexlab{b}})}]{Piz22PCCP}%
  \BibitemOpen
  \bibfield  {author} {\bibinfo {author} {\bibfnamefont {L.}~\bibnamefont
  {Pizzagalli}},\ }\href {https://doi.org/10.1039/d2cp00622g} {\bibfield
  {journal} {\bibinfo  {journal} {Phys. Chem. Chem. Phys.}\ }\textbf {\bibinfo
  {volume} {24}},\ \bibinfo {pages} {9449} (\bibinfo {year}
  {2022}{\natexlab{b}})}\BibitemShut {NoStop}%
\bibitem [{Sup({\natexlab{a}})}]{Sup1}%
  \BibitemOpen
  \href@noop \ \bibinfo {note} {see the dedicated
  section in Supplementary Material, which includes
  Ref.~\cite{Bar04JCP}.}\BibitemShut {Stop}%
\bibitem [{\citenamefont {Giannozzi}\ \emph {et~al.}(2017)\citenamefont
  {Giannozzi}, \citenamefont {Andreussi}, \citenamefont {Brumme}, \citenamefont
  {Bunau}, \citenamefont {Nardelli}, \citenamefont {Calandra}, \citenamefont
  {Car}, \citenamefont {Cavazzoni}, \citenamefont {Ceresoli}, \citenamefont
  {Cococcioni}, \citenamefont {Colonna}, \citenamefont {Carnimeo},
  \citenamefont {Corso}, \citenamefont {de~Gironcoli}, \citenamefont {Delugas},
  \citenamefont {DiStasio}, \citenamefont {Ferretti}, \citenamefont {Floris},
  \citenamefont {Fratesi}, \citenamefont {Fugallo}, \citenamefont {Gebauer},
  \citenamefont {Gerstmann}, \citenamefont {Giustino}, \citenamefont {Gorni},
  \citenamefont {Jia}, \citenamefont {Kawamura}, \citenamefont {Ko},
  \citenamefont {Kokalj}, \citenamefont {Kü{\c{c}}ükbenli}, \citenamefont
  {Lazzeri}, \citenamefont {Marsili}, \citenamefont {Marzari}, \citenamefont
  {Mauri}, \citenamefont {Nguyen}, \citenamefont {Nguyen}, \citenamefont {de-la
  Roza}, \citenamefont {Paulatto}, \citenamefont {Ponc{\'{e}}}, \citenamefont
  {Rocca}, \citenamefont {Sabatini}, \citenamefont {Santra}, \citenamefont
  {Schlipf}, \citenamefont {Seitsonen}, \citenamefont {Smogunov}, \citenamefont
  {Timrov}, \citenamefont {Thonhauser}, \citenamefont {Umari}, \citenamefont
  {Vast}, \citenamefont {Wu},\ and\ \citenamefont {Baroni}}]{Gia17JPCM}%
  \BibitemOpen
  \bibfield  {author} {\bibinfo {author} {\bibfnamefont {P.}~\bibnamefont
  {Giannozzi}}, \bibinfo {author} {\bibfnamefont {O.}~\bibnamefont
  {Andreussi}}, \bibinfo {author} {\bibfnamefont {T.}~\bibnamefont {Brumme}},
  \bibinfo {author} {\bibfnamefont {O.}~\bibnamefont {Bunau}}, \bibinfo
  {author} {\bibfnamefont {M.~B.}\ \bibnamefont {Nardelli}}, \bibinfo {author}
  {\bibfnamefont {M.}~\bibnamefont {Calandra}}, \bibinfo {author}
  {\bibfnamefont {R.}~\bibnamefont {Car}}, \bibinfo {author} {\bibfnamefont
  {C.}~\bibnamefont {Cavazzoni}}, \bibinfo {author} {\bibfnamefont
  {D.}~\bibnamefont {Ceresoli}}, \bibinfo {author} {\bibfnamefont
  {M.}~\bibnamefont {Cococcioni}}, \bibinfo {author} {\bibfnamefont
  {N.}~\bibnamefont {Colonna}}, \bibinfo {author} {\bibfnamefont
  {I.}~\bibnamefont {Carnimeo}}, \bibinfo {author} {\bibfnamefont {A.~D.}\
  \bibnamefont {Corso}}, \bibinfo {author} {\bibfnamefont {S.}~\bibnamefont
  {de~Gironcoli}}, \bibinfo {author} {\bibfnamefont {P.}~\bibnamefont
  {Delugas}}, \bibinfo {author} {\bibfnamefont {R.~A.}\ \bibnamefont
  {DiStasio}}, \bibinfo {author} {\bibfnamefont {A.}~\bibnamefont {Ferretti}},
  \bibinfo {author} {\bibfnamefont {A.}~\bibnamefont {Floris}}, \bibinfo
  {author} {\bibfnamefont {G.}~\bibnamefont {Fratesi}}, \bibinfo {author}
  {\bibfnamefont {G.}~\bibnamefont {Fugallo}}, \bibinfo {author} {\bibfnamefont
  {R.}~\bibnamefont {Gebauer}}, \bibinfo {author} {\bibfnamefont
  {U.}~\bibnamefont {Gerstmann}}, \bibinfo {author} {\bibfnamefont
  {F.}~\bibnamefont {Giustino}}, \bibinfo {author} {\bibfnamefont
  {T.}~\bibnamefont {Gorni}}, \bibinfo {author} {\bibfnamefont
  {J.}~\bibnamefont {Jia}}, \bibinfo {author} {\bibfnamefont {M.}~\bibnamefont
  {Kawamura}}, \bibinfo {author} {\bibfnamefont {H.-Y.}\ \bibnamefont {Ko}},
  \bibinfo {author} {\bibfnamefont {A.}~\bibnamefont {Kokalj}}, \bibinfo
  {author} {\bibfnamefont {E.}~\bibnamefont {Kü{\c{c}}ükbenli}}, \bibinfo
  {author} {\bibfnamefont {M.}~\bibnamefont {Lazzeri}}, \bibinfo {author}
  {\bibfnamefont {M.}~\bibnamefont {Marsili}}, \bibinfo {author} {\bibfnamefont
  {N.}~\bibnamefont {Marzari}}, \bibinfo {author} {\bibfnamefont
  {F.}~\bibnamefont {Mauri}}, \bibinfo {author} {\bibfnamefont {N.~L.}\
  \bibnamefont {Nguyen}}, \bibinfo {author} {\bibfnamefont {H.-V.}\
  \bibnamefont {Nguyen}}, \bibinfo {author} {\bibfnamefont {A.~O.}\
  \bibnamefont {de-la Roza}}, \bibinfo {author} {\bibfnamefont
  {L.}~\bibnamefont {Paulatto}}, \bibinfo {author} {\bibfnamefont
  {S.}~\bibnamefont {Ponc{\'{e}}}}, \bibinfo {author} {\bibfnamefont
  {D.}~\bibnamefont {Rocca}}, \bibinfo {author} {\bibfnamefont
  {R.}~\bibnamefont {Sabatini}}, \bibinfo {author} {\bibfnamefont
  {B.}~\bibnamefont {Santra}}, \bibinfo {author} {\bibfnamefont
  {M.}~\bibnamefont {Schlipf}}, \bibinfo {author} {\bibfnamefont {A.~P.}\
  \bibnamefont {Seitsonen}}, \bibinfo {author} {\bibfnamefont {A.}~\bibnamefont
  {Smogunov}}, \bibinfo {author} {\bibfnamefont {I.}~\bibnamefont {Timrov}},
  \bibinfo {author} {\bibfnamefont {T.}~\bibnamefont {Thonhauser}}, \bibinfo
  {author} {\bibfnamefont {P.}~\bibnamefont {Umari}}, \bibinfo {author}
  {\bibfnamefont {N.}~\bibnamefont {Vast}}, \bibinfo {author} {\bibfnamefont
  {X.}~\bibnamefont {Wu}},\ and\ \bibinfo {author} {\bibfnamefont
  {S.}~\bibnamefont {Baroni}},\ }\href
  {https://doi.org/10.1088/1361-648x/aa8f79} {\bibfield  {journal} {\bibinfo
  {journal} {J. Phys.: Condens. Matter}\ }\textbf {\bibinfo {volume} {29}},\
  \bibinfo {pages} {465901} (\bibinfo {year} {2017})}\BibitemShut {NoStop}%
\bibitem [{\citenamefont {Perdew}\ \emph {et~al.}(1996)\citenamefont {Perdew},
  \citenamefont {Burke},\ and\ \citenamefont {Ernzerhof}}]{Per96PRL}%
  \BibitemOpen
  \bibfield  {author} {\bibinfo {author} {\bibfnamefont {J.~P.}\ \bibnamefont
  {Perdew}}, \bibinfo {author} {\bibfnamefont {K.}~\bibnamefont {Burke}},\ and\
  \bibinfo {author} {\bibfnamefont {M.}~\bibnamefont {Ernzerhof}},\ }\href@noop
  {} {\bibfield  {journal} {\bibinfo  {journal} {Phys. Rev. Lett.}\ }\textbf
  {\bibinfo {volume} {77}},\ \bibinfo {pages} {3865} (\bibinfo {year}
  {1996})}\BibitemShut {NoStop}%
\bibitem [{\citenamefont {Vanderbilt}(1990)}]{Van90PRB}%
  \BibitemOpen
  \bibfield  {author} {\bibinfo {author} {\bibfnamefont {D.}~\bibnamefont
  {Vanderbilt}},\ }\href@noop {} {\bibfield  {journal} {\bibinfo  {journal}
  {Phys. Rev. B}\ }\textbf {\bibinfo {volume} {41}},\ \bibinfo {pages} {7892}
  (\bibinfo {year} {1990})}\BibitemShut {NoStop}%
\bibitem [{Sup({\natexlab{b}})}]{Sup2}%
  \BibitemOpen
  \href@noop \ \bibinfo {note} {see the dedicated
  section in Supplementary Material, which includes
  Refs.~\cite{Ver97PRE,Mol18PCCP}.}\BibitemShut {Stop}%
\bibitem [{Sup({\natexlab{c}})}]{Sup3}%
  \BibitemOpen
  \href@noop \ \bibinfo {note} {see the dedicated
  section in Supplementary Material, which includes
  Refs.~\cite{Hal67PR,Piz21PML,Mil00NANO}.}\BibitemShut {Stop}%
\bibitem [{\citenamefont {Mahler~Larsen}\ \emph {et~al.}(2016)\citenamefont
  {Mahler~Larsen}, \citenamefont {Schmidt},\ and\ \citenamefont
  {Schi{\o}tz}}]{Mah16MSMSE}%
  \BibitemOpen
  \bibfield  {author} {\bibinfo {author} {\bibfnamefont {P.}~\bibnamefont
  {Mahler~Larsen}}, \bibinfo {author} {\bibfnamefont {S.}~\bibnamefont
  {Schmidt}},\ and\ \bibinfo {author} {\bibfnamefont {J.}~\bibnamefont
  {Schi{\o}tz}},\ }\href {https://doi.org/10.1088/0965-0393/24/5/055007}
  {\bibfield  {journal} {\bibinfo  {journal} {Modelling Simul. Mater. Sci.
  Eng.}\ }\textbf {\bibinfo {volume} {24}},\ \bibinfo {pages} {055007}
  (\bibinfo {year} {2016})}\BibitemShut {NoStop}%
\bibitem [{\citenamefont {Rodney}\ \emph {et~al.}(2017)\citenamefont {Rodney},
  \citenamefont {Ventelon}, \citenamefont {Clouet}, \citenamefont
  {Pizzagalli},\ and\ \citenamefont {Willaime}}]{Rod17AM}%
  \BibitemOpen
  \bibfield  {author} {\bibinfo {author} {\bibfnamefont {D.}~\bibnamefont
  {Rodney}}, \bibinfo {author} {\bibfnamefont {L.}~\bibnamefont {Ventelon}},
  \bibinfo {author} {\bibfnamefont {E.}~\bibnamefont {Clouet}}, \bibinfo
  {author} {\bibfnamefont {L.}~\bibnamefont {Pizzagalli}},\ and\ \bibinfo
  {author} {\bibfnamefont {F.}~\bibnamefont {Willaime}},\ }\href
  {https://doi.org/10.1016/j.actamat.2016.09.049} {\bibfield  {journal}
  {\bibinfo  {journal} {Acta Mater.}\ }\textbf {\bibinfo {volume} {124}},\
  \bibinfo {pages} {633} (\bibinfo {year} {2017})}\BibitemShut {NoStop}%
\bibitem [{\citenamefont {Miller}\ and\ \citenamefont
  {Rodney}(2008)}]{Mil08JMPS}%
  \BibitemOpen
  \bibfield  {author} {\bibinfo {author} {\bibfnamefont {R.}~\bibnamefont
  {Miller}}\ and\ \bibinfo {author} {\bibfnamefont {D.}~\bibnamefont
  {Rodney}},\ }\href {https://doi.org/10.1016/j.jmps.2007.10.005} {\bibfield
  {journal} {\bibinfo  {journal} {J. Mech. Phys. Solids}\ }\textbf {\bibinfo
  {volume} {56}},\ \bibinfo {pages} {1203} (\bibinfo {year}
  {2008})}\BibitemShut {NoStop}%
\bibitem [{\citenamefont {Gutkin}\ and\ \citenamefont
  {Ovid'ko}(2008)}]{Gut08AM}%
  \BibitemOpen
  \bibfield  {author} {\bibinfo {author} {\bibfnamefont {M.}~\bibnamefont
  {Gutkin}}\ and\ \bibinfo {author} {\bibfnamefont {I.}~\bibnamefont
  {Ovid'ko}},\ }\href {https://doi.org/10.1016/j.actamat.2007.12.004}
  {\bibfield  {journal} {\bibinfo  {journal} {Acta Mater.}\ }\textbf {\bibinfo
  {volume} {56}},\ \bibinfo {pages} {1642} (\bibinfo {year}
  {2008})}\BibitemShut {NoStop}%
\bibitem [{Sup({\natexlab{d}})}]{Sup4}%
  \BibitemOpen
  \href@noop \ \bibinfo {note} {see the dedicated
  section in Supplementary Material, which includes
  Refs.~\cite{Zen20PRL,Zha11JAP,Hon18AM}.}\BibitemShut {Stop}%
\bibitem [{\citenamefont {Kilymis}\ \emph {et~al.}(2019)\citenamefont
  {Kilymis}, \citenamefont {G{\'{e}}rard},\ and\ \citenamefont
  {Pizzagalli}}]{Kil19AM}%
  \BibitemOpen
  \bibfield  {author} {\bibinfo {author} {\bibfnamefont {D.}~\bibnamefont
  {Kilymis}}, \bibinfo {author} {\bibfnamefont {C.}~\bibnamefont
  {G{\'{e}}rard}},\ and\ \bibinfo {author} {\bibfnamefont {L.}~\bibnamefont
  {Pizzagalli}},\ }\href {https://doi.org/10.1016/j.actamat.2018.11.009}
  {\bibfield  {journal} {\bibinfo  {journal} {Acta Mater.}\ }\textbf {\bibinfo
  {volume} {164}},\ \bibinfo {pages} {560} (\bibinfo {year}
  {2019})}\BibitemShut {NoStop}%
\bibitem [{\citenamefont {Chen}\ \emph {et~al.}(2020)\citenamefont {Chen},
  \citenamefont {Pethö}, \citenamefont {Sologubenko}, \citenamefont {Ma},
  \citenamefont {Michler}, \citenamefont {Spolenak},\ and\ \citenamefont
  {Wheeler}}]{Che20NATCOM}%
  \BibitemOpen
  \bibfield  {author} {\bibinfo {author} {\bibfnamefont {M.}~\bibnamefont
  {Chen}}, \bibinfo {author} {\bibfnamefont {L.}~\bibnamefont {Pethö}},
  \bibinfo {author} {\bibfnamefont {A.~S.}\ \bibnamefont {Sologubenko}},
  \bibinfo {author} {\bibfnamefont {H.}~\bibnamefont {Ma}}, \bibinfo {author}
  {\bibfnamefont {J.}~\bibnamefont {Michler}}, \bibinfo {author} {\bibfnamefont
  {R.}~\bibnamefont {Spolenak}},\ and\ \bibinfo {author} {\bibfnamefont
  {J.~M.}\ \bibnamefont {Wheeler}},\ }\href@noop {} {\bibfield  {journal}
  {\bibinfo  {journal} {Nature Communications}\ }\textbf {\bibinfo {volume}
  {11}} (\bibinfo {year} {2020})}\BibitemShut {NoStop}%
\bibitem [{\citenamefont {He}\ \emph {et~al.}(2016)\citenamefont {He},
  \citenamefont {Fei}, \citenamefont {Tang}, \citenamefont {Zhong},\ and\
  \citenamefont {Meng}}]{He16JNR}%
  \BibitemOpen
  \bibfield  {author} {\bibinfo {author} {\bibfnamefont {Q.}~\bibnamefont
  {He}}, \bibinfo {author} {\bibfnamefont {J.}~\bibnamefont {Fei}}, \bibinfo
  {author} {\bibfnamefont {C.}~\bibnamefont {Tang}}, \bibinfo {author}
  {\bibfnamefont {J.}~\bibnamefont {Zhong}},\ and\ \bibinfo {author}
  {\bibfnamefont {L.}~\bibnamefont {Meng}},\ }\bibfield  {journal} {\bibinfo
  {journal} {J. Nanopart. Res.}\ }\textbf {\bibinfo {volume} {18}},\ \href
  {https://doi.org/10.1007/s11051-016-3358-5} {10.1007/s11051-016-3358-5}
  (\bibinfo {year} {2016})\BibitemShut {NoStop}%
\bibitem [{\citenamefont {Kayang}\ and\ \citenamefont
  {Volkov}(2021)}]{Kay21APA}%
  \BibitemOpen
  \bibfield  {author} {\bibinfo {author} {\bibfnamefont {K.~W.}\ \bibnamefont
  {Kayang}}\ and\ \bibinfo {author} {\bibfnamefont {A.~N.}\ \bibnamefont
  {Volkov}},\ }\href@noop {} {\bibfield  {journal} {\bibinfo  {journal} {Appl.
  Phys.\ A}\ }\textbf {\bibinfo {volume} {127}} (\bibinfo {year}
  {2021})}\BibitemShut {NoStop}%
\bibitem [{\citenamefont {Shin}\ \emph {et~al.}(2012)\citenamefont {Shin},
  \citenamefont {Jin}, \citenamefont {Kim},\ and\ \citenamefont
  {Park}}]{Shi12JACS}%
  \BibitemOpen
  \bibfield  {author} {\bibinfo {author} {\bibfnamefont {C.}~\bibnamefont
  {Shin}}, \bibinfo {author} {\bibfnamefont {H.-H.}\ \bibnamefont {Jin}},
  \bibinfo {author} {\bibfnamefont {W.-J.}\ \bibnamefont {Kim}},\ and\ \bibinfo
  {author} {\bibfnamefont {J.-Y.}\ \bibnamefont {Park}},\ }\href@noop {}
  {\bibfield  {journal} {\bibinfo  {journal} {J. Am. Chem. Soc.}\ }\textbf
  {\bibinfo {volume} {95}},\ \bibinfo {pages} {2944} (\bibinfo {year}
  {2012})}\BibitemShut {NoStop}%
\bibitem [{\citenamefont {Korte}\ and\ \citenamefont {Clegg}(2011)}]{Kor11PM}%
  \BibitemOpen
  \bibfield  {author} {\bibinfo {author} {\bibfnamefont {S.}~\bibnamefont
  {Korte}}\ and\ \bibinfo {author} {\bibfnamefont {W.}~\bibnamefont {Clegg}},\
  }\href {https://doi.org/10.1080/14786435.2010.505179} {\bibfield  {journal}
  {\bibinfo  {journal} {Philos. Mag.}\ }\textbf {\bibinfo {volume} {91}},\
  \bibinfo {pages} {1150} (\bibinfo {year} {2011})}\BibitemShut {NoStop}%
\bibitem [{\citenamefont {Dunstan}\ and\ \citenamefont
  {Bushby}(2013)}]{Dun13IJP}%
  \BibitemOpen
  \bibfield  {author} {\bibinfo {author} {\bibfnamefont {D.}~\bibnamefont
  {Dunstan}}\ and\ \bibinfo {author} {\bibfnamefont {A.}~\bibnamefont
  {Bushby}},\ }\href {https://doi.org/10.1016/j.ijplas.2012.08.002} {\bibfield
  {journal} {\bibinfo  {journal} {Int. J. of Plasticity}\ }\textbf {\bibinfo
  {volume} {40}},\ \bibinfo {pages} {152} (\bibinfo {year} {2013})}\BibitemShut
  {NoStop}%
\bibitem [{\citenamefont {Feruz}\ and\ \citenamefont
  {Mordehai}(2016)}]{Fer16AM}%
  \BibitemOpen
  \bibfield  {author} {\bibinfo {author} {\bibfnamefont {Y.}~\bibnamefont
  {Feruz}}\ and\ \bibinfo {author} {\bibfnamefont {D.}~\bibnamefont
  {Mordehai}},\ }\href
  {https://doi.org/http://dx.doi.org/10.1016/j.actamat.2015.10.027} {\bibfield
  {journal} {\bibinfo  {journal} {Acta Mater.}\ }\textbf {\bibinfo {volume}
  {103}},\ \bibinfo {pages} {433 } (\bibinfo {year} {2016})}\BibitemShut
  {NoStop}%
\bibitem [{\citenamefont {Faisal}\ and\ \citenamefont
  {Weinberger}(2019)}]{Fai19PRM}%
  \BibitemOpen
  \bibfield  {author} {\bibinfo {author} {\bibfnamefont {A.~H.~M.}\
  \bibnamefont {Faisal}}\ and\ \bibinfo {author} {\bibfnamefont {C.~R.}\
  \bibnamefont {Weinberger}},\ }\href
  {https://doi.org/10.1103/physrevmaterials.3.103601} {\bibfield  {journal}
  {\bibinfo  {journal} {Phys. Rev. Materials}\ }\textbf {\bibinfo {volume}
  {3}},\ \bibinfo {pages} {103601} (\bibinfo {year} {2019})}\BibitemShut
  {NoStop}%
\bibitem [{\citenamefont {Roundy}\ and\ \citenamefont
  {Cohen}(2001)}]{Rou01PRB}%
  \BibitemOpen
  \bibfield  {author} {\bibinfo {author} {\bibfnamefont {D.}~\bibnamefont
  {Roundy}}\ and\ \bibinfo {author} {\bibfnamefont {M.}~\bibnamefont {Cohen}},\
  }\href@noop {} {\bibfield  {journal} {\bibinfo  {journal} {Phys. Rev. B}\
  }\textbf {\bibinfo {volume} {64}},\ \bibinfo {pages} {212103} (\bibinfo
  {year} {2001})}\BibitemShut {NoStop}%
\bibitem [{\citenamefont {Dubois}\ \emph {et~al.}(2006)\citenamefont {Dubois},
  \citenamefont {Rignanese}, \citenamefont {Pardoen},\ and\ \citenamefont
  {Charlier}}]{Dub06PRB}%
  \BibitemOpen
  \bibfield  {author} {\bibinfo {author} {\bibfnamefont {S.~M.-M.}\
  \bibnamefont {Dubois}}, \bibinfo {author} {\bibfnamefont {G.-M.}\
  \bibnamefont {Rignanese}}, \bibinfo {author} {\bibfnamefont {T.}~\bibnamefont
  {Pardoen}},\ and\ \bibinfo {author} {\bibfnamefont {J.-C.}\ \bibnamefont
  {Charlier}},\ }\href {https://doi.org/10.1103/PhysRevB.74.235203} {\bibfield
  {journal} {\bibinfo  {journal} {Phys. Rev. B}\ }\textbf {\bibinfo {volume}
  {74}},\ \bibinfo {pages} {235203} (\bibinfo {year} {2006})}\BibitemShut
  {NoStop}%
\bibitem [{\citenamefont {Ogata}\ \emph {et~al.}(2004)\citenamefont {Ogata},
  \citenamefont {Li}, \citenamefont {Hirosaki}, \citenamefont {Shibutani},\
  and\ \citenamefont {Yip}}]{Oga04PRB}%
  \BibitemOpen
  \bibfield  {author} {\bibinfo {author} {\bibfnamefont {S.}~\bibnamefont
  {Ogata}}, \bibinfo {author} {\bibfnamefont {J.}~\bibnamefont {Li}}, \bibinfo
  {author} {\bibfnamefont {N.}~\bibnamefont {Hirosaki}}, \bibinfo {author}
  {\bibfnamefont {Y.}~\bibnamefont {Shibutani}},\ and\ \bibinfo {author}
  {\bibfnamefont {S.}~\bibnamefont {Yip}},\ }\href@noop {} {\bibfield
  {journal} {\bibinfo  {journal} {Phys. Rev. B}\ }\textbf {\bibinfo {volume}
  {70}} (\bibinfo {year} {2004})}\BibitemShut {NoStop}%
\bibitem [{\citenamefont {Umeno}\ and\ \citenamefont
  {\v{C}ern\'y}(2008)}]{Ume08PRB}%
  \BibitemOpen
  \bibfield  {author} {\bibinfo {author} {\bibfnamefont {Y.}~\bibnamefont
  {Umeno}}\ and\ \bibinfo {author} {\bibfnamefont {M.}~\bibnamefont
  {\v{C}ern\'y}},\ }\href@noop {} {\bibfield  {journal} {\bibinfo  {journal}
  {Phys. Rev. B}\ }\textbf {\bibinfo {volume} {77}},\ \bibinfo {pages} {100101
  (R)} (\bibinfo {year} {2008})}\BibitemShut {NoStop}%
\bibitem [{\citenamefont {Barnard}\ and\ \citenamefont
  {Zapol}(2004)}]{Bar04JCP}%
  \BibitemOpen
  \bibfield  {author} {\bibinfo {author} {\bibfnamefont {A.~S.}\ \bibnamefont
  {Barnard}}\ and\ \bibinfo {author} {\bibfnamefont {P.}~\bibnamefont
  {Zapol}},\ }\href {https://doi.org/10.1063/1.1775770} {\bibfield  {journal}
  {\bibinfo  {journal} {J. Chem. Phys.}\ }\textbf {\bibinfo {volume} {121}},\
  \bibinfo {pages} {4276} (\bibinfo {year} {2004})}\BibitemShut {NoStop}%
\bibitem [{\citenamefont {Vergeles}\ \emph {et~al.}(1997)\citenamefont
  {Vergeles}, \citenamefont {Maritan}, \citenamefont {Koplik},\ and\
  \citenamefont {Banavar}}]{Ver97PRE}%
  \BibitemOpen
  \bibfield  {author} {\bibinfo {author} {\bibfnamefont {M.}~\bibnamefont
  {Vergeles}}, \bibinfo {author} {\bibfnamefont {A.}~\bibnamefont {Maritan}},
  \bibinfo {author} {\bibfnamefont {J.}~\bibnamefont {Koplik}},\ and\ \bibinfo
  {author} {\bibfnamefont {J.~R.}\ \bibnamefont {Banavar}},\ }\href
  {https://doi.org/10.1103/PhysRevE.56.2626} {\bibfield  {journal} {\bibinfo
  {journal} {Phys. Rev. E}\ }\textbf {\bibinfo {volume} {56}},\ \bibinfo
  {pages} {2626} (\bibinfo {year} {1997})}\BibitemShut {NoStop}%
\bibitem [{\citenamefont {Molleman}\ and\ \citenamefont
  {Hiemstra}(2018)}]{Mol18PCCP}%
  \BibitemOpen
  \bibfield  {author} {\bibinfo {author} {\bibfnamefont {B.}~\bibnamefont
  {Molleman}}\ and\ \bibinfo {author} {\bibfnamefont {T.}~\bibnamefont
  {Hiemstra}},\ }\href {https://doi.org/10.1039/c8cp02346h} {\bibfield
  {journal} {\bibinfo  {journal} {Phys. Chem. Chem. Phys.}\ }\textbf {\bibinfo
  {volume} {20}},\ \bibinfo {pages} {20575} (\bibinfo {year}
  {2018})}\BibitemShut {NoStop}%
\bibitem [{\citenamefont {Hall}(1967)}]{Hal67PR}%
  \BibitemOpen
  \bibfield  {author} {\bibinfo {author} {\bibfnamefont {J.~J.}\ \bibnamefont
  {Hall}},\ }\href@noop {} {\bibfield  {journal} {\bibinfo  {journal} {Phys.
  Rev.}\ }\textbf {\bibinfo {volume} {161}},\ \bibinfo {pages} {756} (\bibinfo
  {year} {1967})}\BibitemShut {NoStop}%
\bibitem [{\citenamefont {Pizzagalli}(2021)}]{Piz21PML}%
  \BibitemOpen
  \bibfield  {author} {\bibinfo {author} {\bibfnamefont {L.}~\bibnamefont
  {Pizzagalli}},\ }\href {https://doi.org/10.1080/09500839.2021.1909167}
  {\bibfield  {journal} {\bibinfo  {journal} {Philos. Mag. Lett.}\ }\textbf
  {\bibinfo {volume} {101}},\ \bibinfo {pages} {242} (\bibinfo {year}
  {2021})}\BibitemShut {NoStop}%
\bibitem [{\citenamefont {Miller}\ and\ \citenamefont
  {Shenoy}(2000)}]{Mil00NANO}%
  \BibitemOpen
  \bibfield  {author} {\bibinfo {author} {\bibfnamefont {R.}~\bibnamefont
  {Miller}}\ and\ \bibinfo {author} {\bibfnamefont {V.}~\bibnamefont
  {Shenoy}},\ }\href@noop {} {\bibfield  {journal} {\bibinfo  {journal}
  {Nanotechnology}\ }\textbf {\bibinfo {volume} {11}},\ \bibinfo {pages} {139}
  (\bibinfo {year} {2000})}\BibitemShut {NoStop}%
\bibitem [{\citenamefont {Zeng}\ \emph {et~al.}(2020)\citenamefont {Zeng},
  \citenamefont {Zeng}, \citenamefont {Ge}, \citenamefont {Chen}, \citenamefont
  {Lou}, \citenamefont {Chen}, \citenamefont {Yan}, \citenamefont {Yang},
  \citenamefont {kwang Mao}, \citenamefont {Yang},\ and\ \citenamefont
  {Mao}}]{Zen20PRL}%
  \BibitemOpen
  \bibfield  {author} {\bibinfo {author} {\bibfnamefont {Z.}~\bibnamefont
  {Zeng}}, \bibinfo {author} {\bibfnamefont {Q.}~\bibnamefont {Zeng}}, \bibinfo
  {author} {\bibfnamefont {M.}~\bibnamefont {Ge}}, \bibinfo {author}
  {\bibfnamefont {B.}~\bibnamefont {Chen}}, \bibinfo {author} {\bibfnamefont
  {H.}~\bibnamefont {Lou}}, \bibinfo {author} {\bibfnamefont {X.}~\bibnamefont
  {Chen}}, \bibinfo {author} {\bibfnamefont {J.}~\bibnamefont {Yan}}, \bibinfo
  {author} {\bibfnamefont {W.}~\bibnamefont {Yang}}, \bibinfo {author}
  {\bibfnamefont {H.}~\bibnamefont {kwang Mao}}, \bibinfo {author}
  {\bibfnamefont {D.}~\bibnamefont {Yang}},\ and\ \bibinfo {author}
  {\bibfnamefont {W.~L.}\ \bibnamefont {Mao}},\ }\bibfield  {journal} {\bibinfo
   {journal} {Phys. Rev. Lett.}\ }\textbf {\bibinfo {volume} {124}},\ \href
  {https://doi.org/10.1103/physrevlett.124.185701}
  {10.1103/physrevlett.124.185701} (\bibinfo {year} {2020})\BibitemShut
  {NoStop}%
\bibitem [{\citenamefont {Zhang}\ \emph {et~al.}(2011)\citenamefont {Zhang},
  \citenamefont {Deng}, \citenamefont {Hong}, \citenamefont {Xiong},
  \citenamefont {Li}, \citenamefont {Strasberg}, \citenamefont {Yin},
  \citenamefont {Zou}, \citenamefont {Taylor}, \citenamefont {Sawyer},\ and\
  \citenamefont {Chen}}]{Zha11JAP}%
  \BibitemOpen
  \bibfield  {author} {\bibinfo {author} {\bibfnamefont {N.}~\bibnamefont
  {Zhang}}, \bibinfo {author} {\bibfnamefont {Q.}~\bibnamefont {Deng}},
  \bibinfo {author} {\bibfnamefont {Y.}~\bibnamefont {Hong}}, \bibinfo {author}
  {\bibfnamefont {L.}~\bibnamefont {Xiong}}, \bibinfo {author} {\bibfnamefont
  {S.}~\bibnamefont {Li}}, \bibinfo {author} {\bibfnamefont {M.}~\bibnamefont
  {Strasberg}}, \bibinfo {author} {\bibfnamefont {W.}~\bibnamefont {Yin}},
  \bibinfo {author} {\bibfnamefont {Y.}~\bibnamefont {Zou}}, \bibinfo {author}
  {\bibfnamefont {C.~R.}\ \bibnamefont {Taylor}}, \bibinfo {author}
  {\bibfnamefont {G.}~\bibnamefont {Sawyer}},\ and\ \bibinfo {author}
  {\bibfnamefont {Y.}~\bibnamefont {Chen}},\ }\href
  {https://doi.org/http://dx.doi.org/10.1063/1.3552985} {\bibfield  {journal}
  {\bibinfo  {journal} {J. Appl. Phys.}\ }\textbf {\bibinfo {volume} {109}},\
  \bibinfo {pages} {063534} (\bibinfo {year} {2011})}\BibitemShut {NoStop}%
\bibitem [{\citenamefont {Hong}\ \emph {et~al.}(2018)\citenamefont {Hong},
  \citenamefont {Zhang},\ and\ \citenamefont {Zaeem}}]{Hon18AM}%
  \BibitemOpen
  \bibfield  {author} {\bibinfo {author} {\bibfnamefont {Y.}~\bibnamefont
  {Hong}}, \bibinfo {author} {\bibfnamefont {N.}~\bibnamefont {Zhang}},\ and\
  \bibinfo {author} {\bibfnamefont {M.~A.}\ \bibnamefont {Zaeem}},\ }\href
  {https://doi.org/https://doi.org/10.1016/j.actamat.2017.11.034} {\bibfield
  {journal} {\bibinfo  {journal} {Acta Mater.}\ }\textbf {\bibinfo {volume}
  {145}},\ \bibinfo {pages} {8 } (\bibinfo {year} {2018})}\BibitemShut
  {NoStop}%
\end{thebibliography}

%

\end{document}